\documentclass[a4paper]{article}
\usepackage{indentfirst}
\usepackage{hyperref}
\usepackage{multirow}
\usepackage{latexsym,bm,amsmath,amssymb}
\usepackage{graphicx}
\usepackage{epsfig}
\usepackage{subfigure}
\usepackage{cite}
\usepackage{color}
\usepackage[top=1in, bottom=1in, left=1.25in, right=1.25in]{geometry}
\usepackage{slashed}
\usepackage{fancyhdr}
\usepackage{slashed}
\usepackage{makecell,rotating,multirow,diagbox}
\usepackage{tikz}
\usepackage{float}
\usepackage{appendix}
\usepackage{setspace}
\usepackage{geometry}
\usepackage{graphics}
\usepackage{multirow}
\usepackage{booktabs}
\usetikzlibrary{shapes.geometric, arrows}
\usetikzlibrary{trees}
\usetikzlibrary{decorations.pathmorphing}
\usetikzlibrary{decorations.markings}
\tikzset{
        photon/.style={decorate, decoration={snake}, draw=red},
        nucleon/.style={draw=black, postaction={decorate},
           decoration={markings,mark=at position .55 with{\arrow[draw=black]{>}}}},
        pion/.style={draw=blue, postaction={decorate},
        decoration={markings,mark=at position .55 with{\arrow[draw=blue]{}}}},
        sigma/.style={draw=black, postaction={decorate},
        decoration={markings,mark=at position .55 with{\arrow[draw=black]{}}}},
        link/.style    = { draw=black, double = white, line width = 1.8pt, double distance = 0.8pt , postaction={decorate},decoration={markings,mark=at position .55 with{\arrow[draw=black]{>}}}},
    }

\newcommand{\be}{\begin{equation}}
\newcommand{\ee}{\end{equation}}
\newcommand{\ba}{\begin{array}{c}} \newcommand{\ea}{\end{array}}
\newcommand{\bqa}{\begin{eqnarray}}
\newcommand{\eqa}{\end{eqnarray}}

\def\bea{\arraycolsep .1em \begin{eqnarray}}
\def\eea{\end{eqnarray}}

\def\s0#1#2{\mbox{\small{$ \frac{#1}{#2} $}}}
\def\0#1#2{\frac{#1}{#2}}

\begin{document}
\setcounter{topnumber}{10}
\setcounter{totalnumber}{50}
\newcommand{\red}[1]{ {\color{red} #1}}
\newcommand{\blue}[1]{ {\color{blue} #1}}

\title{Some remarks on compositeness of $T^+_{cc}$}
\maketitle
\begin{center}
{\sc
Chang Chen$^{\dagger\,}$,\,
Ce Meng$^{\dagger}$,\,
Zhiguang Xiao$^{\heartsuit,}$\footnote{Corresponding author.},\,
Han-Qing~Zheng$^{\heartsuit}$
}
\\
\vspace{0.5cm}
\noindent{\small{$^\dagger$ \it  Department of Physics,
 Peking University, Beijing 100871, P.~R.~China}}\\
\noindent{\small{$^\heartsuit$ \it  College of Physics, Sichuan University, Chengdu, Sichuan 610065, P.~R.~China}}\\
\end{center}

\begin{abstract}
Recently, the LHCb experimental group found an exotic state $T^+_{cc}$ from
the $pp \to D^0D^0\pi^+ + X$ process. A key question is whether it is just
a molecule or may have confined tetraquark ingredient. To investigate
this, different methods were used, including a two-channel
($D^{*+}D^0$ and $D^{*0}D^+$) $K$-matrix unitarization and a 
single-channel Flatt\'e-like parametrization method analyzed 
utilizing the pole counting
rule and spectral density function sum rule. These analyses
demonstrated that $T^+_{cc}$ is a molecular state, although the
possibility that there may exist an elementary ingredient cannot be
excluded, according to an approximate analysis of its production rate.
\end{abstract}

\section{Introduction}

The LHCb collaboration found a very narrow peak structure named
$T^+_{cc}$ in the $D^0D^0\pi^+$ invariant mass spectrum, in the $pp \to X + D^0D^0\pi^+ $ process\cite{LHCb:2021vvq}.
The mass parameters obtained from a generic constant-width
Breit-Wigner fit were listed as
$$
\delta m_{BW} = -273 \pm 61 \pm 5^{+11}_{-14}~ \mathrm{keV}\ ,\,\,\,
\Gamma_{BW} = 410 \pm 165 \pm 43^{+18}_{-38}~ \mathrm{keV}\,,
$$
where $\delta m_{BW}$ defines the mass shift with respect to the $D^{*+}D^0$ threshold.
Later it was suggested that $T^+_{cc}$ could more possibly be an isoscalar
state with spin-parity quantum numbers $J^P = 1^+$\cite{LHCb:2021auc},
and in a more complicated model, the pole mass and 
width were extracted as
$$
\delta m_{pole} = -360 \pm 40^{+4}_{-0} ~ \mathrm{keV}\ ,\,\,\,
\Gamma_{pole} = 48\pm 2^{+0}_{-14}~ \mathrm{keV}\,.
$$ 
The constituent of $T^+_{cc}$ is $cc\bar{u}\bar{d}$ and there is no
annihilated quark pair, similar to $X_1(2900)$ ($ud\bar{s}\bar{c}$)\cite{LHCb:2020pxc, Chen:2021tad}.

This experimental observation has stimulated numerous theoretical
discussions. First of all, there have been some dynamical lattice QCD
simulations about double charmed tetraquarks, although they have not
provided a definite conclusion on the existence of the $T^+_{cc}$ state\cite{Ikeda:2013vwa,
Junnarkar:2018twb}. Recently, based on (2 + 1)-flavor lattice QCD
simulations, $D^*D$ system was studied more carefully. It was verified
that there is a loosely-bound state near the threshold (-10 keV)
\cite{Lyu:2023xro}. Many phenomenological studies have also been
conducted.
A theoretical prediction is that there may exist a $cc\bar{u}\bar{d}$
tetraquark with $J^P = 1^+$ near $D^{*+}D^0$ threshold \cite{Karliner:2017qjm,
Qin:2020zlg}.
In addition, the heavy meson chiral effective field theory (HMChEFT)
 which considers contact and one pion exchange (OPE) interaction was
used.
The prefered conclusion of the  analyses is that $T^+_{cc}$ state is a
molecule of $D^{*+}D^0$ and $D^{*0}D^+$\cite{Du:2021zzh,Lin:2022wmj}.
The effect
of triangle diagram singularity was also evaluated with $D^*D\pi$
interactions. It was found that the contribution is very weak compared
with that of the tree diagram, which 
suggests that $T^+_{cc}$ is not generated from the triangle singularity~\cite{Achasov:2022onn}.
The pole parameters of $T^+_{cc}$ extracted from a simple $K$-matrix amplitude
were also studied and it was found that $T^+_{cc}$ may originate from a $D^{*+}D^0$ virtual state~\cite{Dai:2021wxi}. 
The extended chiral 
Lagrangian with $K$-matrix unitarity approach was also applied, and it
was suggested that vector meson exchanges could play a crucial role in forming $T^+_{cc}$ bound state of $D^*D$~\cite{Feijoo:2021ppq}. 

In this work, to determine whether $T^+_{cc}$ is just a
loosely-bound $s$-wave molecule of $D^*D$ or it contains the
$cc\bar{u}\bar{d}$ ingredient, different approaches are used. First,
we adopte an approach similar to that of Ref.~\cite{Feijoo:2021ppq},
 in which a coupled channel unitarity approach ($D^{*+}D^0$ and $D^{*0}D^+$) 
with the interaction stemming from the extended hidden local gauge
lagrangian is also applied. In Ref.~\cite{Feijoo:2021ppq}, the authors only
considered the vector exchanging diagram contributions and there was no
fitting of the lineshape data. In this study, more complete
interactions,
including pseudoscalar, vector exchanges, and $D^*D$ contact terms, are
introduced, and a combined fit of $DD\pi$, $D\pi$, and $DD$ channels is
made. It indicates that the $\rho$ vector meson exchange couplings
really make non-negligible contributions in generating the $T^+_{cc}$
resonance compared with the other two interactions. In this scheme, there
exists a bound state near the $D^{*+}D^0$ threshold, which suggests
that $T^+_{cc}$ may be a $D^*D$ molecule. Furthermore, the
Flatt\'e-like parametrization is also examined. Through a combined fit on
three-body and two-body invariant mass spectrum, we find that the
result is the same based on the pole counting rule (PCR) and spectral density function sum rule
calculation~\cite{Lu:2023ccs, Chen:2022ddj, Gong:2016hlt, Meng:2014ota,
Zhang:2009bv, Yao:2020bxx}: there is only one pole near $D^*D$
threshold and the corresponding $\mathcal{Z} \simeq 1$. We also
attempt to
judge the compositeness of $T^{+}_{cc}$ by comparing its production
($pp\to T^{+}_{cc} + X$) with different theoretical estimations.
However, it is
 difficult to make a clear judgement using this approach.

This paper is organized as follows: Sec.~I is the introduction,
 in Sec. II, the $K$-matrix unitarization approach using an
effective Lagrangian in $s$-wave approximation is introduced and its numerical fit is shown.
 In Sec.~III, other frameworks are employed to analyze the
compositeness of $T^+_{cc}$. Finally, in Sec.~IV, a brief conclusion
on the structure of $T^+_{cc}$ is drawn.

\section{K-matrix unitarity approach}

We start off from a $SU(4)$ invariant effective Lagrangian, and then
modify the relevant parameters to only preserve the $SU(2)$ symmetry
later, as in~\cite{Lin:1999ad}. 
Here, we list the relevant coupling terms
\begin{equation}\label{L}
    \begin{aligned}
    \mathcal{L} = \mathcal{L}_0 - ig\mathrm{Tr}([P,\partial_\mu P]V^\mu)+ig\mathrm{Tr}([V^\nu, \partial_\mu V_\nu]V^\mu)\\
    -\frac{g^2}{2}\mathrm{Tr}([P,V_\mu]^2)+\frac{g^2}{4}\mathrm{Tr}([V_\mu,V_\nu]^2)\ ,
    \end{aligned}
\end{equation}
where $\mathcal{L}_0$ is the free Lagrangian for the pseudoscalar and
vector mesons. $P$ and $V$ denote, respectively, the properly normalized pseudoscalar and vector meson matrices 
\begin{equation}\label{P}
    P = \begin{pmatrix}
        \frac{\eta}{\sqrt{3}}+\frac{\eta'}{\sqrt{6}}+\frac{\pi^0}{\sqrt{2}}&\pi^+&K^+&\bar{D}^0\\
        \pi^-&\frac{\eta}{\sqrt{3}}+\frac{\eta'}{\sqrt{6}}-\frac{\pi^0}{\sqrt{2}}&K^0&D^-\\
        K^-&\bar{K}^0&-\frac{\eta}{\sqrt{3}}+\sqrt{\frac{2}{3}}\eta'&D^-_s\\
        D^0&D^+&D^+_s&\eta_c
    \end{pmatrix}\ ,
\end{equation}
\begin{equation}\label{V}
    V = \begin{pmatrix}
        \frac{\omega}{\sqrt{2}}+\frac{\rho^0}{\sqrt{2}}&\rho^+&K^{*+}&\bar{D}^{*0}\\
        \rho^-&\frac{\omega}{\sqrt{2}}-\frac{\rho^0}{\sqrt{2}}&K^{*0}&D^{*-}\\
        K^{*-}&\bar{K}^{*0}&\phi&D^{*-}_s\\
        D^{*0}&D^{*+}&D^{*+}_s&J/\psi
    \end{pmatrix}\ .
\end{equation}
In the later discussions, all the coupling constants denoted as $g$ in
the vertices of the PPV, VVV and PPVV types in Eqs.  (\ref{P}) and (\ref{V})
could be different for the vertices of different isospin multiplets,
such that only the $SU(2)$ isospin symmetry is retained.  Hereafter, we only consider
the vertices relevant to our discussions. We adopt
previous theoretical works~\cite{Meng:2014ota,Yao:2015qia,Lin:1999ad}
to estimate these coupling constants because the interaction vertices
are similar to theirs with only normalization constant differences,
which can be tracked with a careful analysis. 

\begin{figure}[h]
    \centering
    \subfigure[]{
    \includegraphics[scale = 0.3]{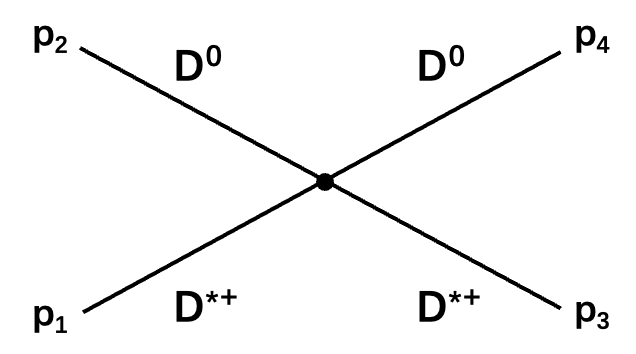}
    }
    \subfigure[]{
    \includegraphics[scale = 0.3]{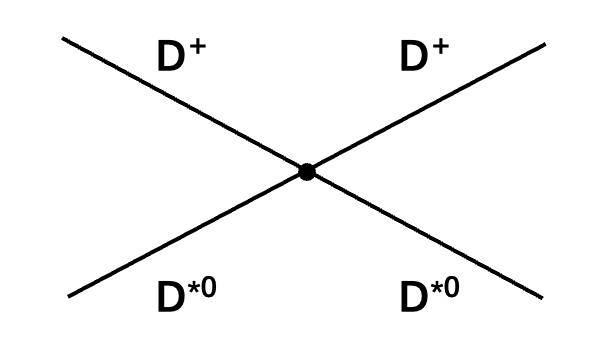}
    }
    \subfigure[]{
    \includegraphics[scale = 0.3]{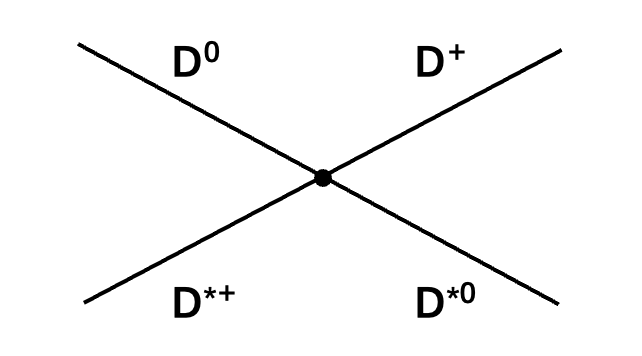}
    }
    \caption{Contact diagrams.\label{fig:contact}}
    \label{contact}
\end{figure}

From Eq.~(\ref{L}), we can obtain the contact, $t$ and $u$ channel
diagrams of the $D^{*+}D^0\to D^{*0}D^+$ process. We list their
amplitudes successively. First, for the contact diagrams
in Fig. \ref{fig:contact},
\begin{equation}\label{Mc}
iM^{c}_{ij} = i~g_{D^*DD^*D}^2,
\end{equation}
where $i,j = 1,2$ refer to the $D^{*+}D^0$ and $D^{*0}D^+$ channel,
respectively. The  coupling $g_{D^*DD^*D}$ was estimated when studying
$X(3872)$~\cite{Meng:2014ota},  that is, $g (=g_{D^*DD^*D}) \simeq
16$\footnote{In Ref.~\cite{Meng:2014ota}, all short range $D^*D$
interaction effect is reflected in the contact coupling strength
instead of separating them as contact and vector meson
exchange terms like what we do here. There may be some risk
of the double counting to take this specific value. However, since 
these contact terms contribute just a  
smooth background, it would not affect the analysis of the sharp peak in
our numerical analysis.}.

\begin{figure}[h]
    \centering
    \includegraphics[scale = 0.3]{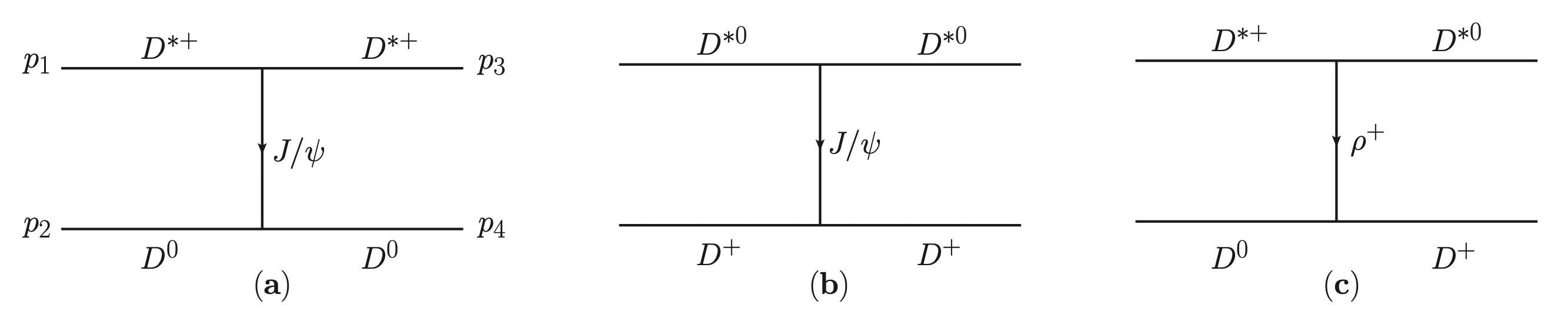}\\
    \caption{$T$ channel diagrams.}
    \label{t}
\end{figure}

The $t$ channel diagrams include vector meson ($J/\psi$ or $\rho$,
$\omega$) exchanges~\cite{Feijoo:2021ppq}, as shown in
Fig.~\ref{t}.\footnote{The $\omega$ and
$\rho^0$ exchange diagram have the same coupling constant with
opposite signs. So they almost cancel each other.} We also neglect the
momentum dependence in the denominator of the  propagator  near the
threshold. Here, the estimates about the coupling  constants from the PPV
and VVV vertices are  as follows: For $i, j = 1,1$ or $2,2$, the
coupling constant $g(= g_{J/\psi D^{(*)}D^{(*)}})\simeq 7.7$, and for
$i, j = 1,2$ or $2,1$, $g(= g_{\rho D^{(*)}D^{(*)}})\simeq 3.9$, which
 are obtained from the vector meson dominance (VMD)
assumption~\cite{Lin:1999ad}. 
The $t$-channel amplitudes are hence written down as follows:
\begin{equation}\label{Mt}
    iM^t_{ij} = i D_{ij} (p_1 +p_3)\cdot(p_2 +p_4)~\epsilon(p_1)\cdot\epsilon^*(p_3),
\end{equation}
where 
\begin{equation}
D_{ij} = \begin{pmatrix}
    \frac{g^2_{J/\psi D^{(*)}D^{(*)}}}{M^2_{J/\psi}}&\frac{g^2_{\rho D^{(*)}D^{(*)}}}{m^2_\rho}\\
    \frac{g^2_{\rho D^{(*)}D^{(*)}}}{m^2_\rho}&\frac{g^2_{J/\psi D^{(*)}D^{(*)}}}{M^2_{J/\psi}}\\
\end{pmatrix}_{ij}.
\end{equation}

\begin{figure}[h]
    \centering
    \subfigure[]{
    \includegraphics[scale = 0.3]{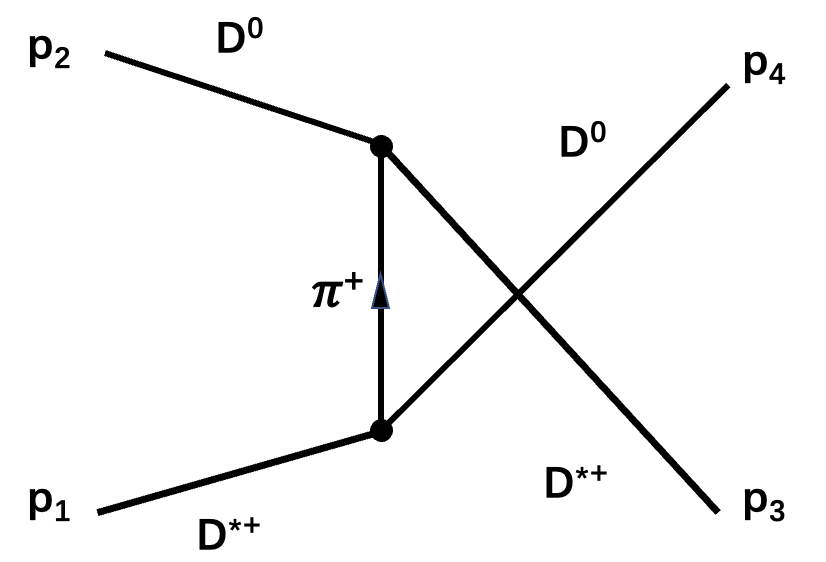}
    }
    \subfigure[]{
    \includegraphics[scale = 0.3]{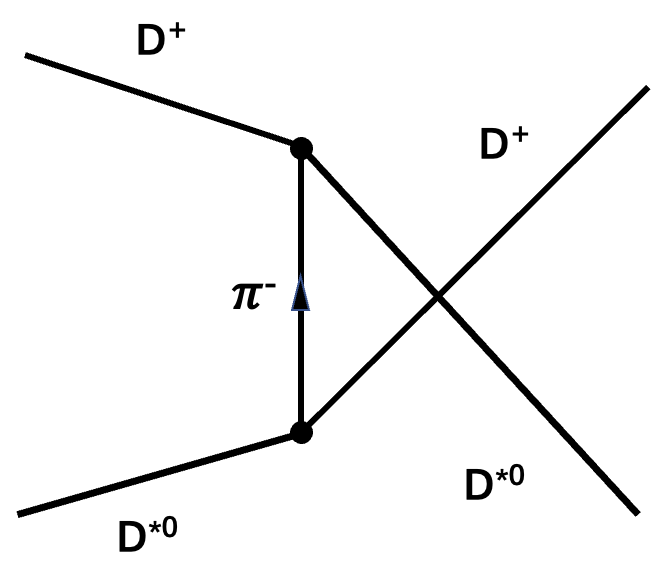}
    }
    \subfigure[]{
    \includegraphics[scale = 0.3]{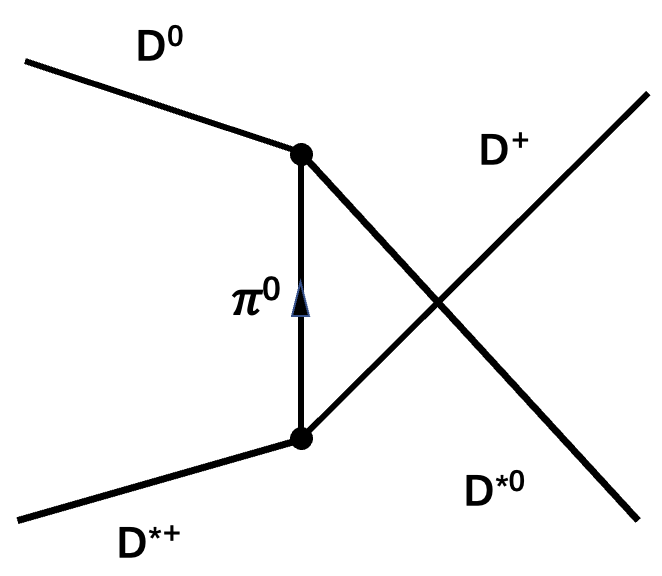}
    }
    \caption{$U$ channel diagrams.}
    \label{u}
\end{figure}

The third type are $u$ channel diagrams with $\pi$ exchanges as shown in Fig.~\ref{u},
and the amplitudes are 
\begin{equation}\label{Mu}
    iM^u_{ij} = i E_{ij} g_{\pi DD^*}^2\frac{\epsilon(p_1)\cdot (p_1-2p_4)~\epsilon^*(p_3)\cdot (p_3-2p_2)}{(p_1-p_4)^2-m_\pi^2},
\end{equation}
where 
\begin{equation}
E_{ij} = \begin{pmatrix}
    -1&\frac{1}{2}\\
    \frac{1}{2}&-1\\
\end{pmatrix}_{ij}.
\end{equation}
The coupling strength $g_{\pi DD^*}$ can be restricted by the decay
process $D^{*+} \to D^0 \pi^+$, and we take the value $g(=g_{\pi
DD^*}) \simeq 8.4$~\cite{Yao:2015qia}. $U$-channel
diagrams with pseudo-scalar meson $\eta_c$ exchanges also exist. They are not
important at this energy range as it is tested numerically, so we neglect these diagrams.
As for the amplitudes corresponding to Fig.~\ref{u}, the $u$-channel
$\pi$ exchange is somewhat special because it is possible to
exchange one on-shell $\pi$ meson. After partial wave projection,
there exists, in  tree level amplitudes,  an additional  cut
in the energy region above the $D^{*+}D^0$ threshold.
Here, this singularity will affect the unitarity.
To remedy this, we adopt the approximate relation $m_{D^*} =
m_{D} + m_{\pi}$ to keep the unitarity threshold  away from the singularity, as in Refs.~\cite{Du:2021zzh,Lin:2022wmj}.
At last, we obtain the total coupled channel amplitudes
\begin{equation}
    M_{ij} = M^c_{ij} + M^t_{ij} +M^u_{ij} \ .
\end{equation}

Furthermore, with the assumption that the full amplitude is mainly
contributed by the $s$-wave amplitude and the $d$-wave amplitude can be
neglected, we consider the $s$-wave amplitude, which can be unitarized by the relation 
    \begin{equation}\label{G}
         \mathbf{T}^{-1} = \mathbf{K}^{-1} -  \mathbf{g}(s),
    \end{equation}
    where $\mathbf{T}$
    is the unitarized $s$-wave scattering $T$ matrix, $\mathbf{K}$  is the  
two-channel $s$-wave scattering amplitude matrix from $M_{ij}$~\cite{Chung:186421}, and $\mathbf{g}(s)\equiv
\mathrm{diag}\{g_i(s)\}$. In our normalization convention
    \begin{equation}\label{gis}
        g_i(s;M_i,m_i) = -16\pi^2 i\int \frac{d^4q}{(2\pi)^4} \frac{1}{(q^2-M_i^2+i\epsilon)((P-q)^2-m_i^2+i\epsilon)},\,\, (s = P^2)
    \end{equation}
    where $M_i$ is the vector meson mass and $m_i$ is the pseudoscalar meson mass in the $i$-th channel.
    The expression of $g_i(s)$ in Eq.~(\ref{gis}) is renormalized using the standard $\overline{\mathrm{MS}}$ scheme, which introduces an explicit renormalization scale ($\mu$) dependence.  In our fit, we select the same $\mu$ parameter in the two channels.

    To obtain a finite width for the $T_{cc}^+$ state below the $D^*D$ threshold, we need to consider the finite width of the $D^*$ state. This is accomplished by performing a convolution of the $g_i(s)$ functions with the mass distribution of the $D^*$ states~\cite{Wang:2019mph}:
    \begin{equation}
        S\left(s_V;M_V,\Gamma_V\right) \equiv -\frac{1}{\pi} \operatorname{Im}\left\{\frac{1}{s_V-M_V^2+i M_V \Gamma_V}\right\}
    \end{equation}
    {such that}
    \begin{equation}\label{gauss}
        \tilde{g}_i(s;M_i,m_i)=\mathcal{C} \int_{s_{Vmin}}^{s_{Vmax}} d s_V g_i(s;\sqrt{s_V},m_i) S\left(s_V;M_i,\Gamma_{i}\right)\ ,
    \end{equation}
    {where $\mathcal{C}$ is a normalization factor. }
    {The main contribution to this integration is from the region
around the unstable mass  $s_V\sim M_V^2$, so we can introduce a
cutoff $s_{Vmin}$ and $s_{Vmax}$. For example, for $\tilde{g}_1$, it is
integrated from $(m_{D_0}+m_{\pi^+})^2$ to
$(m_{D^{*+}}+2\Gamma_{D^{*+}})^2$, whereas for $\tilde{g}_2$, it is
integrated from $(m_{D_0}+m_{\pi^0})^2$ to
$(m_{D^{*0}}+2\Gamma_{D^{*0}})^2$.} Here, in principle, $\Gamma_i$
is $s$-dependent as in Ref.~\cite{Feijoo:2021ppq}. However, we
approximate the decay widths as
constants because we only focus on a small region near the $D^{*+}D^0$
threshold, and we also verified that it makes little difference
 if the $s$ dependence
is included in the numerical calculations. The constant decay widths suggested by PDG~\cite{ParticleDataGroup:2020ssz} read
    \begin{equation}
        \Gamma_{D^{*+}} = 83.4~\mathrm{keV},~~\Gamma_{D^{*0}} = 55.3~\mathrm{keV}.
    \end{equation} 

\begin{figure}[h]
    \centering
    \includegraphics[scale = 0.4]{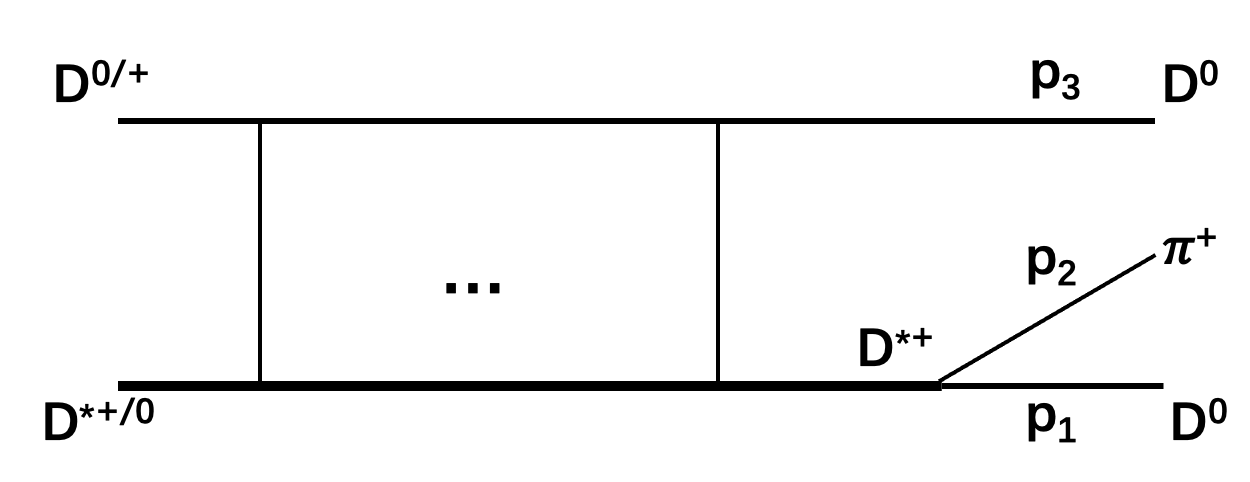}\\
    \caption{Process of $T^+_{cc}\to D^0D^0\pi^+$.}
    \label{santi}
\end{figure}

To fit the final state three-body invariant mass spectrum  of
$D^0D^0\pi^+$ in $pp \to X + D^0D^0\pi^+ $, the  final-state
interaction (FSI)~\cite{Yao:2020bxx} between  $D^{*+}D^0$ and/or
$D^{*0}D^+$ needs to be considered as above, before contemplating the
$D^{*+}\to D^0+\pi^+$ decay.  The amplitude for the $D^{*+} D^0$ final state reads
\begin{equation}
    \mathcal{F}_{D^{*+} D^{0}}(s)=\alpha_{1}~T_{11}+\alpha_{2}~T_{21}\ ,
\end{equation}
where
$\alpha_1$, $\alpha_2$ are smooth real polynomials parametrizing the
amplitude of producing $D^{*+}D^0$ and $D^{*0}D^+$, respectively, and as the energy region of interest is very small, we treat them as constant parameters near the thresholds of $D^{*+}D^0$ and $D^{*0}D^+$.
Finally, the decay of $T^+_{cc}\to D^0D^0\pi^+$ can therefore be
expressed as in Fig.~\ref{santi}, and the final $s$-wave scattering amplitude
is written as\footnote{Analogous equation is used in
\cite{Feijoo:2021ppq} earlier. The difference is that here the
propagator of $D^*$ is written in unitary gauge rather than Feynman
gauge. However, these two gauges make little numerical difference near and below $D^{*+}D^0$ threshold region.}

\begin{equation}\label{gamma}
    \begin{aligned}
        t = &\mathcal{F}_{D^{*+}D^0}\left[
        \frac{ {\epsilon}\cdot[({p}_1-{p}_2)+\frac{m_{\pi^+}^2-m_{D^0}^2}{m_{D^{*+}}^2}({p}_1+{p}_2)
        ]}{M_{12}^2-m_{D^{*+}}^2+iM_{12}\Gamma_{D^{*+}}(M_{12})}\right.\\
        &\left.+\frac{ {\epsilon}\cdot [({p}_3-{p}_2)+\frac{m_{\pi^+}^2-m_{D^0}^2}{m_{D^{*+}}^2}({p}_3+{p}_2)
        ]}{M_{23}^2-m_{D^{*+}}^2+iM_{23}\Gamma_{D^{*+}}(M_{23})}
    \right].
    \end{aligned}
\end{equation}
Here, $M_{12}$ and $M_{23}$ are Dalitz kinematic variables of the
final three-body state. The corresponding definition is $s_{ij}=M_{ij}^2 = (p_i + p_j)^2$, and  ${\epsilon}={\epsilon}(P)$ corresponds to the polarization vector of $T^+_{cc}$, $P=\left(p_{1}+p_{2}+p_{3}\right)$, and $P^{2}=s$. These invariants have the relation $M^2_{12} + M^2_{13} + M^2_{23} = P^2 + p_1^2+ p_2^2+ p_3^2$. 
Finally, the decay width of $T_{cc}^+$ is given by
\begin{equation}\label{123}
    d\Gamma(\sqrt{s}) = \frac{\mathcal{N}}{2} \frac{32}{\pi}\frac{1}{s^{3/2}} |t|^2 ds_{12}ds_{23}.
\end{equation}
The factor $\frac{1}{2}$ in the above equation results from averaging
the two integrals of $D^0$ in the final state. To fit the
experimental data, the normalization factor $\mathcal{N}$ should be
introduced. As for the two FSI parameters, $\alpha_1$ can be absorbed
in the coefficients $\mathcal{N}$. Thus, $\alpha_1 = 1$ is fixed and
there remains one free parameter $\alpha_2$. Besides, 
to obtain the yields for the $D^0D^0\pi^+$ invariant mass
spectrum, the resolution function needs to be  convoluted
\begin{equation}
   \mathrm{Yields}(l)=\int_{l-2 \sigma}^{l+2 \sigma} d l^{\prime} \frac{1}{\sqrt{2 \pi} \sigma} \Gamma\left(l^{\prime}\right) \mathrm{e} ^{-\frac{\left(l^{\prime}-l\right)^2}{2 \sigma^2}},
\end{equation}
where $\sigma = 1.05 \times 263 \mathrm{keV}$~\cite{LHCb:2021vvq}.
At last, invariant mass distributions for the selected two-body state (particles 2 and  3, for example) can also be derived as the following function:
\begin{equation}\label{21}
\frac{d Br}{dM_{23}} = \mathcal{N}'
\int_{m^2_{D^0D^0\pi^+}}^{m_{max}^2} ds \int ds_{12}
|t(s,s_{12},s_{23})|^2
\end{equation}
where $\mathcal{N}'$ is another normalization constant, and $M_{23}$ is
the invariant mass of particles 2 and  3. The  $T^+_{cc}$ energy is
integrated from the initial energy $m_{D^0D^0\pi^+}$ to a cutoff
$m_{max}$.\footnote{Since $T_{cc}^+$  lies just below the threshold of
$D^*D$ with a sharp peak, we can take a rough cutoff about one or two
times its Breit–Wigner widths above the threshold. The subsequent results are not sensitive to this uncertainty.}

Data obtained from the LHCb collaboration about three-body final states
$D^0D^0\pi^+$\cite{LHCb:2021vvq} and two-body invariant mass
distributions $D^0\pi^+$, $D^0D^0$, and $D^+D^0$\cite{LHCb:2021auc} are
used to make a combined fit. The normalization $\mathcal{N}$,
$\mathcal{N'}$, FSI parameter $\alpha_2$, and renormalization scale
$\mu$ are regarded as free parameters to be fitted and all coupling
constants found in the literature are regarded as fixed parameters.

\begin{figure}[h]
        \centering
        \includegraphics[scale = 0.6]{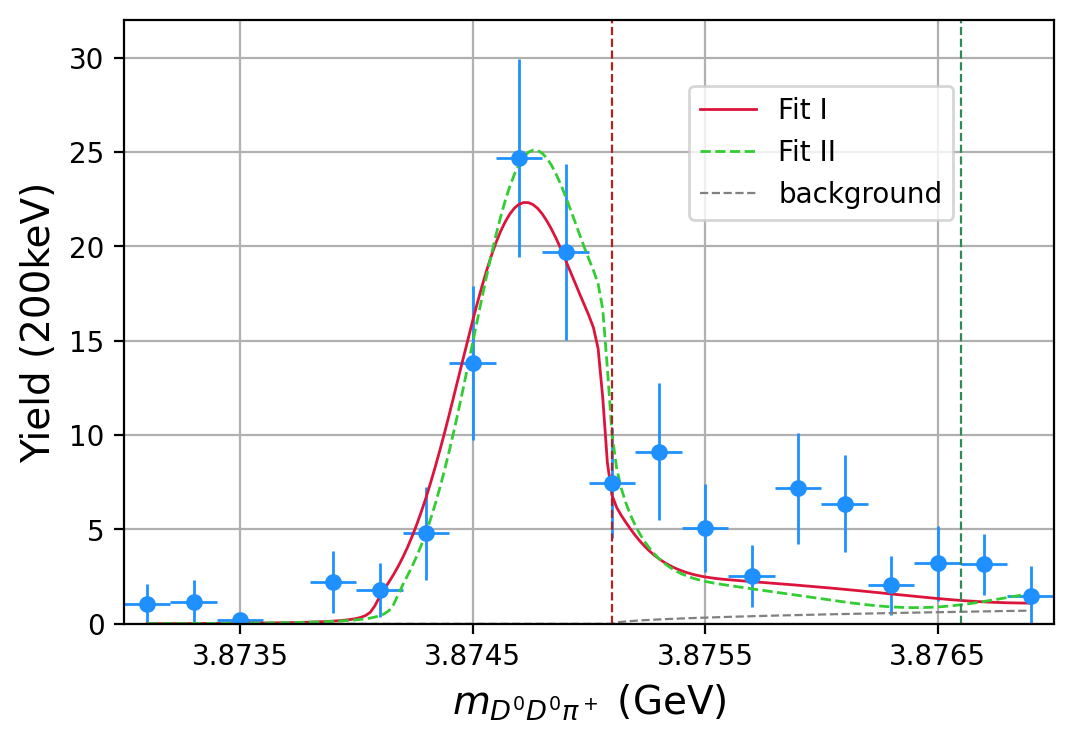}\\
        \caption{$D^0D^0\pi^+$ final state invariant mass spectrum. The vertical purple dash line indicates the 
        $D^{*+}D^0$ threshold and the green one corresponds to the $D^{*0}D^+$ threshold. Data obtained from Ref.~\cite{LHCb:2021vvq}.}\label{Fig5}
\end{figure}

The fit result is presented in Fig.~\ref{Fig5} and Table~\ref{tab3}. 
It is found that the fit
result is very sensitive to the parameter $\mu$. That occurs because the
peak ($T^+_{cc}$ state) is too narrow, considering that the unit of $\mu$
is GeV but the signal range is in MeV. The discussion above seems to
suggest that the fit result prefers a particular choice of parameter
$\mu$. In \cite{Feijoo:2021ppq}, a specific fixed $\mu=1.5$GeV was also
taken in their analysis, which is similar to our result.

The pole location on the $s$-plane is also studied. 
If $D^{*}$ is taken as a stable particle, then $T_{cc}^+$ appears as  a bound state pole located at $\sqrt{s} = 3.8746$, that is, approximately $500 $keV
below the $D^{*+}D^0$ threshold ($\sqrt{s} = 3.8751 $). As there is no accompanying virtual pole nearby, we conclude that, according to the pole counting rule, $T^+_{cc}$ is a pure molecule composed of $DD^*$. 
However, due to the instability of $D^*$, the  $D^{*}D$ channel opens at the energy somewhat smaller than  
$m_{D^*} + m_D$ and the decay of $T_{cc}^+$ takes place~\cite{Feijoo:2021ppq}.

\begin{figure}[H]
    \centering
    \includegraphics[scale = 0.6]{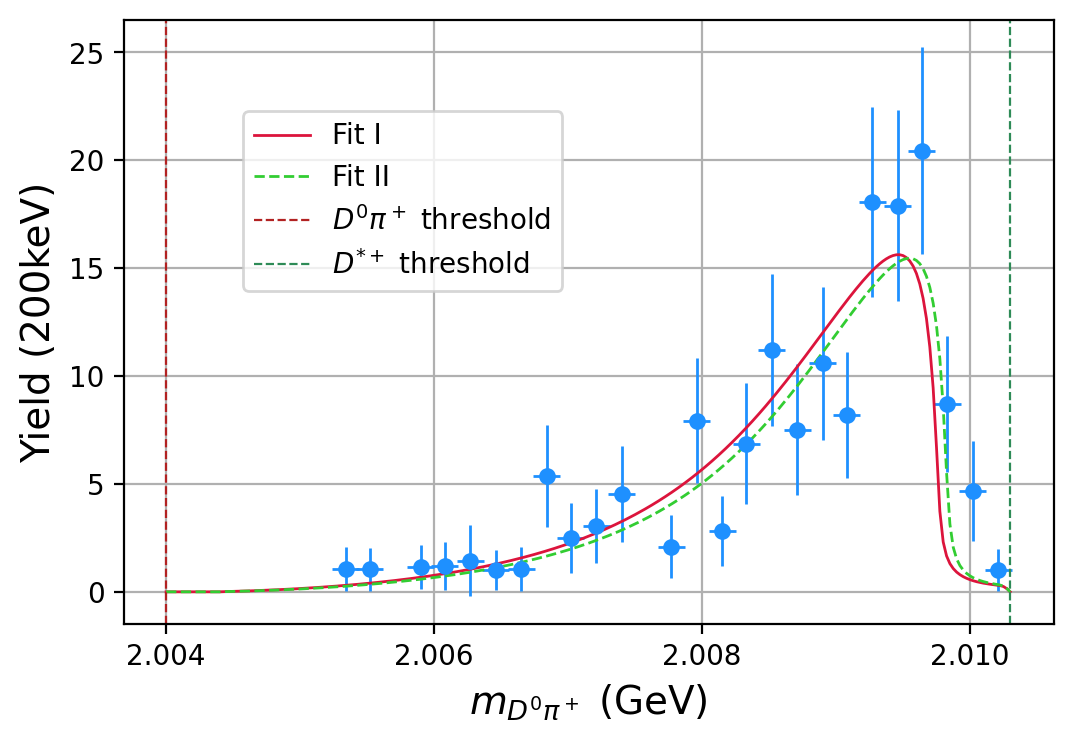}\\
    \caption{$D^0\pi^+$ invariant mass spectrum from the three body final state $D^0D^0+X$ (Data from Ref.~\cite{LHCb:2021auc}).\label{D0pi}}
\end{figure}

\begin{figure}[H]
    \centering
    \subfigure[]{
    \includegraphics[scale = 0.5]{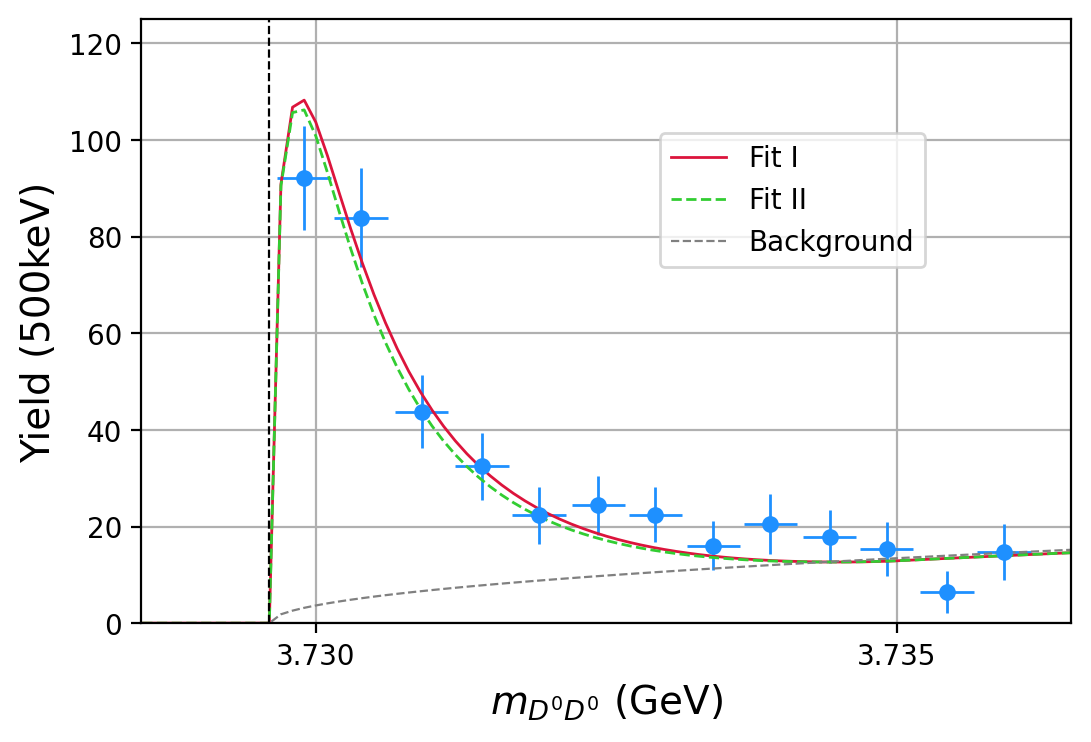}
    }
    \subfigure[]{
    \includegraphics[scale = 0.5]{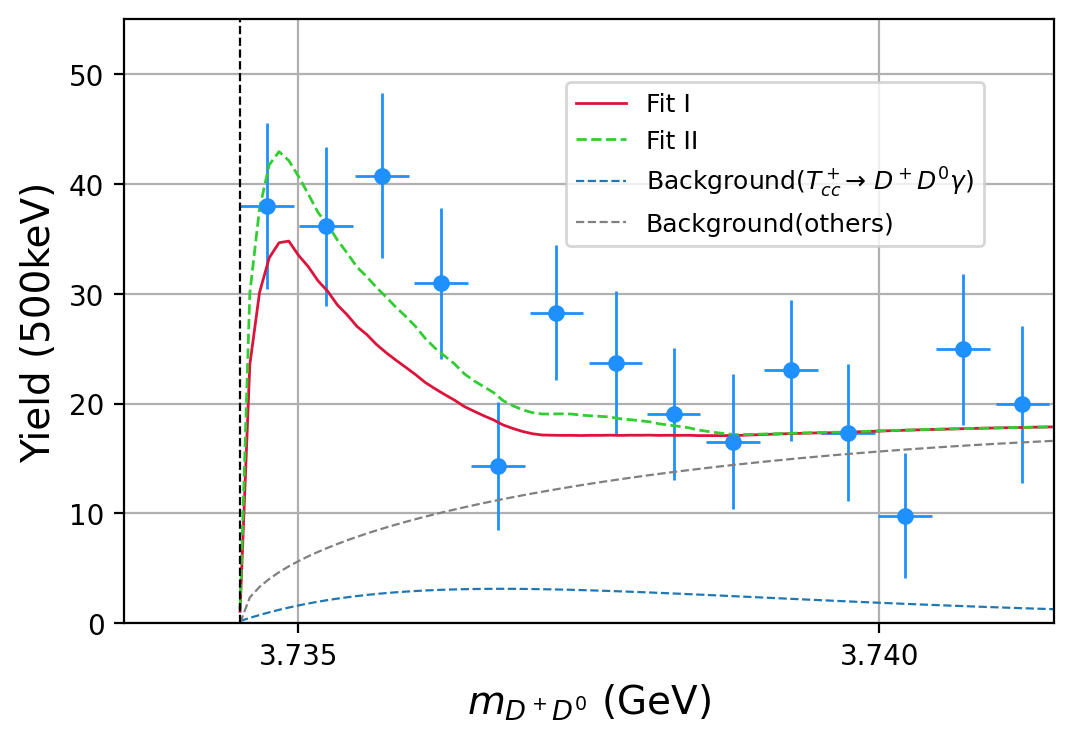}
    }
    \caption{$D^0D^0$ ($D^+D^0$) invariant mass spectrum from the three body final state $D^0D^0+X$ ($D^+D^0 + X$) \cite{LHCb:2021auc}, where the vertical dashed line indicates the $D^{0}D^{0}$ ($D^+D^0$) threshold.}\label{2FS}
\end{figure}

Furthermore, invariant mass distributions for any two of three final
state particles are also taken into consideration. As for the $D^0\pi^+$
state, which comes from  $D^{*+}D^0$, we take $m_{max} =
3.8751\mathrm{GeV}$.  The fit result is shown in Fig.
\ref{D0pi}.
As for $D^0D^0$ states, we take the same $m_{max} = 3.8751\mathrm{GeV}$  ($\mathcal{N}' = \mathcal{N}_{DD}$ here).
The  $D^+D^0$ final state, which comes from the $D^+D^0\pi^0$ final state, is  different. Since  $D^+D^0$ state may come from two  channels, $D^{*+}D^0$ and $D^{*0}D^+$, they need to be considered altogether aided by isospin symmetry. 
Since the threshold of the second channel is higher, we take $m_{max} = 3.8766\mathrm{GeV}$, and on account of a symmetry factor $\frac{1}{2}$ in the channel including $D^0D^0$, the normalization constant here is doubled
($\mathcal{N}' = 2\mathcal{N}_{DD}$). The fitting results are plotted in Fig.~\ref{2FS}. 
Both invariant mass spectra ($D^0D^0$and $D^+D^0$)
have an incoherent background component, parametrized as a product of the two-body phase space function $\Phi_{DD}$ and a linear function. For $D^+D^0$ from channel $T^+_{cc}\to D^{*0}D^+\to D^+D^0\pi^0/D^+D^0\gamma$, because the decay channel $D^{*0}\to D^0\gamma$ accounts for 35\% of the total $D^{*0}$ decay width, this incoherent background contribution is non-negligible and needs to be counted specially. Here, we take this estimation from Ref.~\cite{LHCb:2021auc} directly.

 Finally, the fit parameters are listed in Table \ref{tab3}.  

\begin{table}[h]  
        \caption{Parameters.}\label{tab3}
        \centering
        \begin{tabular}{lc}   
        \toprule
    $\chi^2/d.o.f$ &1.16 \\
    \midrule   
   $\alpha_2$ &  $-0.43\pm 0.10$ \\
    $\mu/\mathrm{GeV}$ &  $1.122\pm 0.001$\\
    $g_{D^*DD^*D}$  & fixed=$16$ \cite{Meng:2014ota}\\
    $g_{\pi DD^*}$  & fixed=$8.4$ \cite{Yao:2015qia}\\
    $g_{\rho D^{(*)}D^{(*)}}$  & fixed=$3.9$ \cite{Lin:1999ad} \\
    $g_{J/\psi D^{(*)}D^{(*)}}$  & fixed=$7.7$ \cite{Lin:1999ad}\\
        \bottomrule  
        \end{tabular}
\end{table}

\section{Other insights on $T^+_{cc}$}

\begin{figure}[H]
    \centering
    \includegraphics[scale = 0.2]{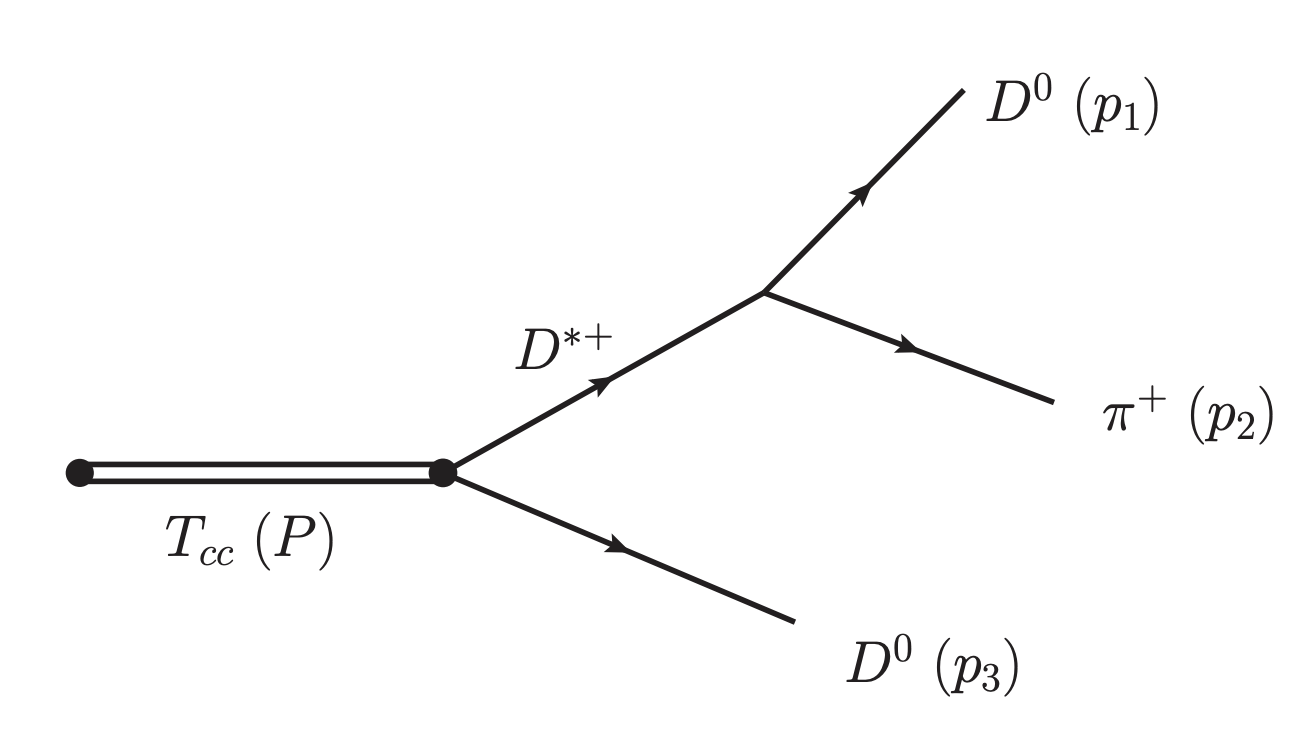}
    \caption{Cascade decay.\label{fig:cascade}}
\end{figure}

In this section, the production of $T^+_{cc}$ in some other methods
is also analyzed to determine its compositeness. First of all, a
single channel Flatt\'e-like parametrization is used, where we do
not distinguish $D^{*+}D^0$ and $D^{*0}D^+$ anymore. As in the
previous calculation in Sec. II, this process is regarded as a cascade
decay, as shown in Fig.~\ref{fig:cascade}. The propagator of $T^+_{cc}$ is approximated by Flatt\'e
formula. The later propagator of $D^*$ selected here is a simple Breit–Wigner amplitude form because the energy
of this process is near the threshold of $D^*D$ and its range is sufficiently small
enough. Besides, the momentum dependent polynomial in the numerator is
also normalized by a constant factor $\mathcal{N}$ for convenience.
Numerical calculations indicate that these approximations  make
little difference. 

\begin{figure}[H]
    \centering
    \subfigure[]{
    \includegraphics[scale = 0.46]{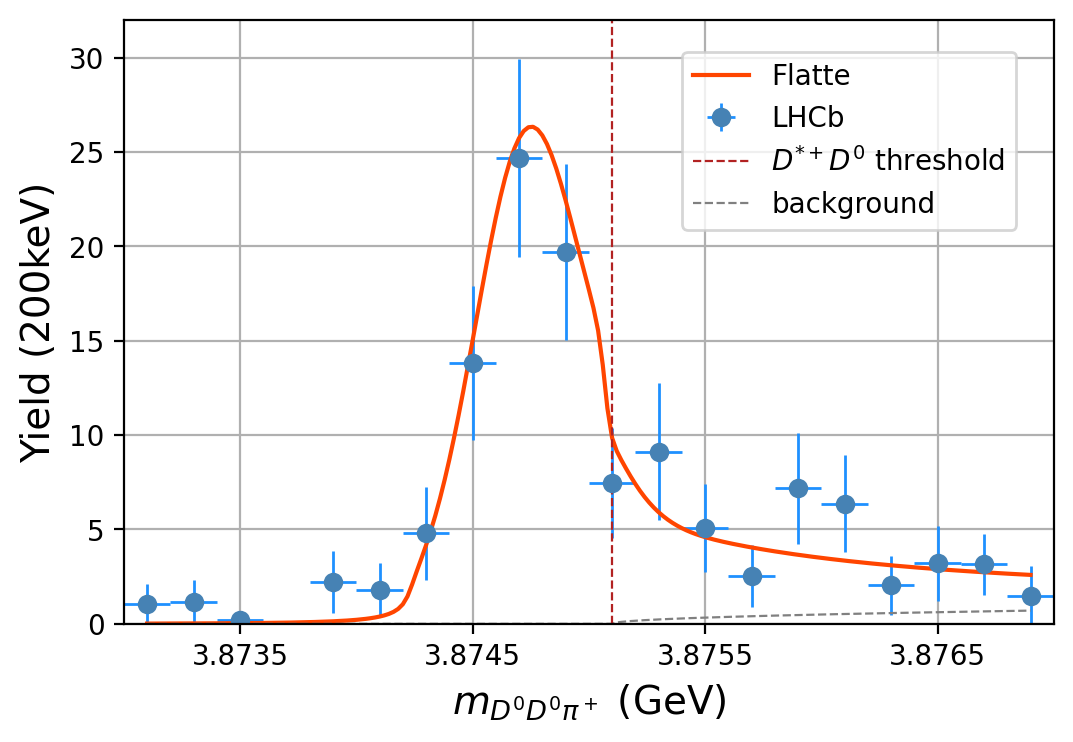}
    }
    \subfigure[]{
    \includegraphics[scale = 0.46]{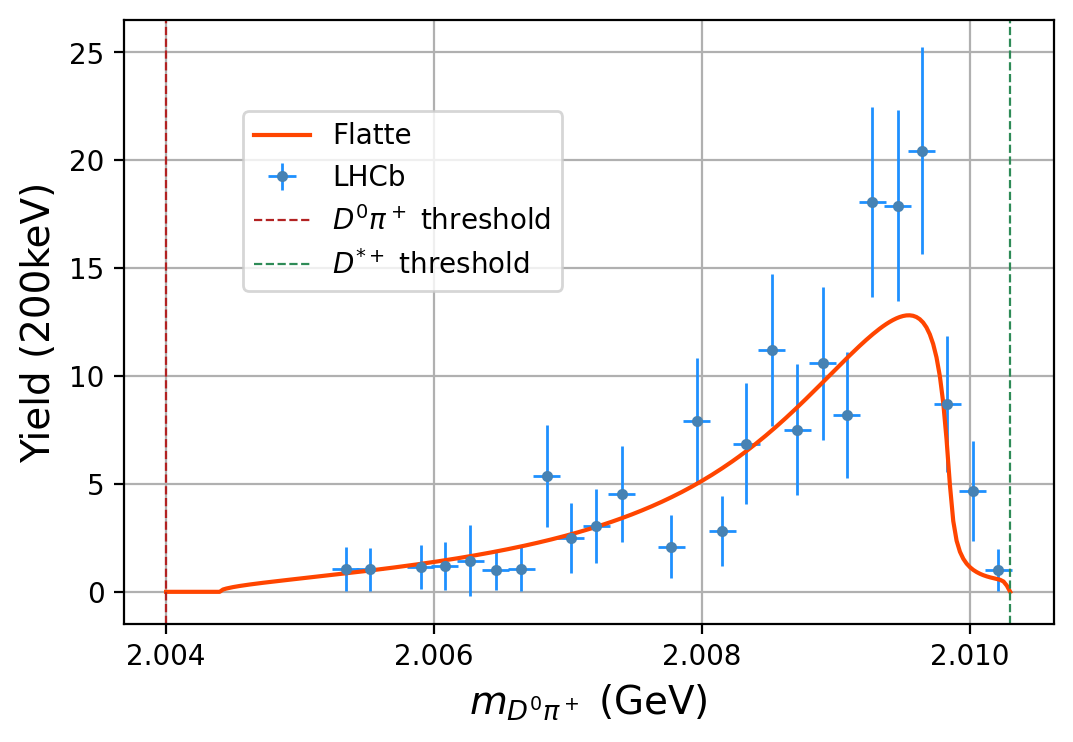}
    }
    \caption{Three-body and two-body invariant mass
spectrum\label{fig:inv-mass-fit}}
\end{figure}

The total $s$-wave approximation amplitude about the process $T^+_{cc} \to D^0D^0\pi^+$ can be written as

\begin{equation}\label{flatte}
    \begin{aligned}
    t= & \frac{1}{s-M^2+i M\left(\hat g \rho(s)\right)}\times\\
    &\left(\frac{1}{M_{12}^2-m_{D^{*+}}^2+i M_{12} \Gamma_{D^{*+}}}+\frac{1}{M_{23}^2-m_{D^{*+}}^2+i M_{23} \Gamma_{D^{*+}}}\right), 
    \end{aligned}
\end{equation}
where $\hat g$ presents the coupling strength with
$D^*D$. The doubling of the kinetic variables of the $D^*$ propagator
is due to the indistinguishability of $D\pi$ in the three-body final
state, and the symmetry factor $\frac{1}{2}$ is absorbed by the total
normalization $\mathcal{N}$. By this parametrization, we make the
energy resolution convolution as that before and fit the three-body decay
width and two-body invariant mass spectra at the same time using
the previous Eqs.~(\ref{123}) and (\ref{21}). It is worth pointing out that
under normal conditions, it will form a divergent peak because of the
zero partial decay width. However, if we regard $D^*$ as an unstable
particle, in other words, if the amplitude can develop an
imaginary part when the energy has not yet reached the $D^*D$ threshold, 
the peak is not divergent anymore. We can use the same stratagem as
Eq. (\ref{gauss}) to treat $\rho$ in Eq. (\ref{flatte}), or more
simply take the value $m_{D^*}$ in $\rho$ with an imaginary part
$\Gamma_{D^*}$. This selection does not affect the result except
for the goodness of fit. Here, we take the latter scenario. The
results of the combined fit are shown in  Fig.~\ref{fig:inv-mass-fit}.

\begin{figure}[h]
    \centering
    \includegraphics[scale = 0.46]{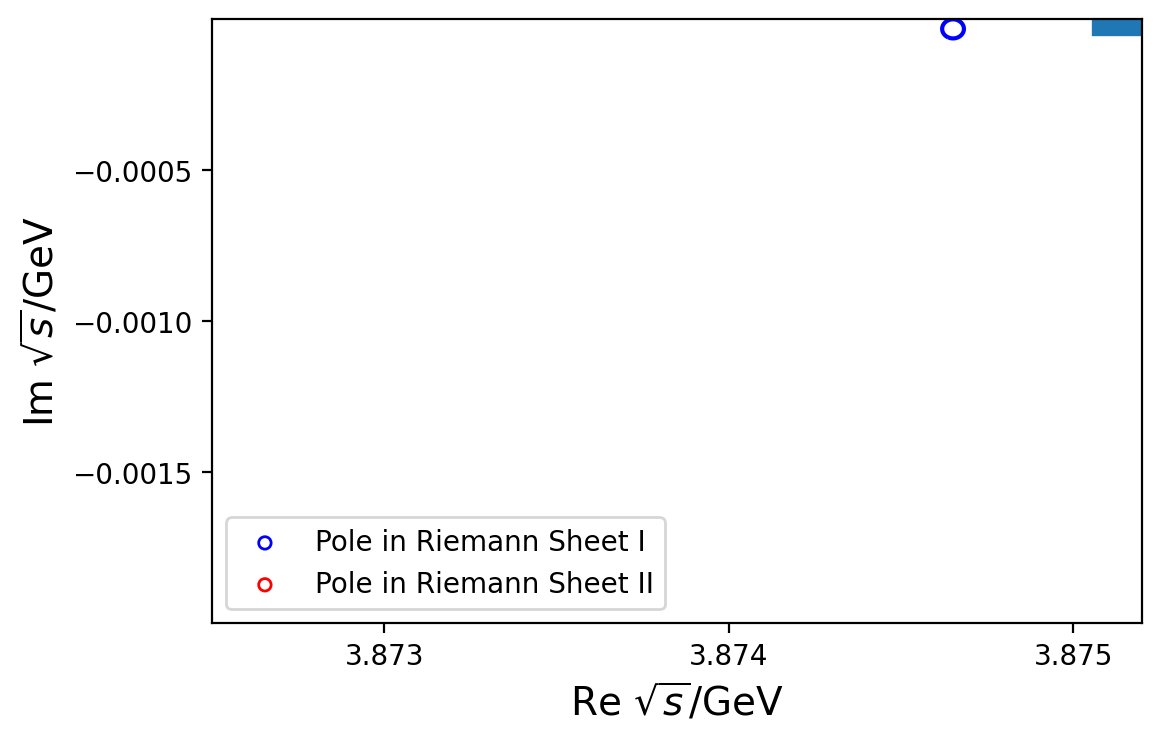}
    \caption{\label{pole stru}Pole structure of flatt\'e amplitude}
\end{figure}

We list the corresponding parameters in Table~\ref{lab5}, and the pole structure of the Flatt\'e amplitude is displayed in Fig.~\ref{pole stru}.
\begin{table}[H]
    \caption{Fit parameter\label{lab5}}
    \centering
    \begin{tabular}{lcc} 
    \toprule   
    $\chi^2/d.o.f $  &$\hat g$ &$M$ \\
    \midrule   
    $0.81$ keV& $ 0.075\pm 0.015$ & $3874.1\pm 0.2 $MeV\\
    \bottomrule
    \end{tabular}
\end{table}

Furthermore, according to the Flatt\'e-like parametrization, it is natural to calculate the probability of finding an `elementary' state in the continuous spectrum by the spectral density function\cite{Baru_2004}
\begin{equation}
    \omega(E)=\frac{1}{2 \pi} \frac{\tilde{g} \sqrt{2 \tilde{M} E} \theta(E)+\tilde{\Gamma}_0}{\left|E-E_f+\frac{\mathrm{i}}{2} \tilde{g} \sqrt{2 \tilde{M} E} +\frac{\mathrm{i}}{2} \tilde{\Gamma}_0\right|^2},
\end{equation}
where $E = \sqrt{s} - m_{D^*D}$, $E_f = M - m_{D^*D}$, $\tilde{M}$ is
the reduced mass of $D^*D$, $\theta$ is the step function at
threshold,
$\tilde{g} = 2 \hat{g} / m_{D^*D}$ is the dimensionless coupling constant of the concerned coupling mode and $\tilde{\Gamma}_0$ is the constant partial width for the remaining couplings. By integrating it with a cut off (usually comparable to the total decay width$\sim \Gamma$), the possibility of finding an `elementary' state in the final state is
\begin{equation}
    \mathcal{Z}=\int_{E_{\min }}^{E_{\max }} \omega(E) \mathrm{d} E.
\end{equation}
Considering that no other channels are coupling with $T^+_{cc}$ under the $D^*D$ threshold, the $\tilde{\Gamma}_0$ here should be set to zero.  In this case, the integrated results in different sections are as follows.

\begin{table}[H]\label{Table3}
    \caption{Spectral density function integrating $\mathcal{Z}$}
    \centering
    \begin{tabular}{lccc} 
    \toprule   
    $[M-\Gamma, M+\Gamma]$ &$[M-2\Gamma, M+2\Gamma]$ &$[M-3\Gamma, M+3\Gamma]$\\
    \midrule   
    $\sim$ 0 & $\sim$ 0 &0.01\\
    \bottomrule
    \end{tabular}
\end{table}
The result suggests that in a simple single channel Flatt\'e-like parametrization framework, $T^+_{cc}$ is a pure molecular state. 
This is in agreement with the result of Ref.~\cite{Dai:2023cyo}, obtained using an effective range expansion approximation. Our result is much more definite than that obtained in Ref.~\cite{LHCb:2021auc}.

Furthermore, the compositeness may also be discussed from the
production rate of a particle. The cross section relation
between the confined states $\Xi_{cc}$($ccu/ccd$)\cite{LHCb:2017iph} and
$T^+_{cc}$ can be obtained from the experiment. 
Since 2016, these two sets of experimental data are both collected 
under the same experimental condition, such as transverse
momentum truncation $p_T$ and luminance $9fb^{-1}$. After taking
account of the detection efficiency and branching fraction
differences\cite{Yu:2017zst}, there is an approximate relation 
\begin{equation}
    \frac{\sigma(pp\to T^+_{cc})}{\sigma(pp\to \Xi_{cc})} \sim  \frac{1}{3} \times \frac{1}{10}.
\end{equation}
If we agree that there exists a universal ratio between the $(Q/QQ)q$ and $(Q/QQ)qq$
productivities in high energy collision\cite{CMS:2012wje}, where $Q$
represents a heavy quark and $q$ is a light quark, we can obtain a factor
$1/3$, which means that catching two light quarks is always more
difficult, that is, $\sigma(pp\to \Xi_{cc}) \simeq 3 \sigma(pp\to
(cc\bar{u}\bar{d}))$. Thus, the ratio between the  cross sections of the observed $T^+_{cc}$ and
the hypothetical tetraquark can be obtained as
\begin{equation}\label{10}
    \frac{\sigma(pp\to T^+_{cc})}{\sigma(pp\to (cc\bar{u}\bar{d}))} \sim  \frac{1}{10}.
\end{equation}
It is also possible to estimate the different orders of
magnitude of the theoretical cross sections between the `elementary' and
`molecular' picture of $T^+_{cc}$. Because the $X(3872)$
resonance has analogous characteristics~\cite{BaBar:2007cmo,
Belle:2008fma}(e.g., binding energy and quark composition),
some comparisons have been made about these orders of magnitude for
$X(3872)$\cite{Braaten:2004ai,Bignamini:2009sk}. If one can borrow the
discussions here, it can be estimated that approximately for $T^+_{cc}$
\begin{equation}\label{100}
    \frac{\sigma(pp\to (c\bar{u})(c\bar{d}))}{\sigma(pp\to (cc\bar{u}\bar{d}))} \sim  \mathcal{O}(10^{-2})-\mathcal{O}(10^{-3}).
\end{equation}
A similar result was also obtained in \cite{Jin:2021cxj}.
By comparing Eq. (\ref{10}) and Eq. (\ref{100}), one can find that
the production of $T^+_{cc}$ just falls in between two different
cases. However, this analysis depends on some uncertain
assumptions and is not quite reliable. In Ref.~\cite{Hua:2023zpa}, the production
cross section for $T_{cc}$ as a molecule was estimated to be approximately an
order of magnitude higher than that as a tetraquack, which creates more
confusion.  Thus, using the production argument cannot provide a clear
conclusion on the nature of $T^+_{cc}$. On the contrary, the analysis
provided in this paper, for example in Table~\ref{Table3}, clearly indicates the
molecular nature of $T^+_{cc}$. 
        
\section{Summary}

In this work, we study the nature of $T^+_{cc}$ using different
methods. First, an effective field theory Lagrangian with two
channels ($D^{*+}D^0$ and $D^{*0}D^+$) combined with a $K$-matrix approach is used  to
describe the $T^+_{cc}\to D^0D^0\pi^+$ process. The three-body and
two-body invariant mass spectrum can be fitted well at the same time.
The numerical fit results reveal that vector meson exchanges
are more important than $\pi$ exchanges and contact
interactions. Second, the Flatt\`e formula is used to
study the same problem. Both approaches suggest  that $T^+_{cc}$ is
definitely a pure molecular state composed of $D^*D$, in agreement
with many of the results found in the literature, but on a much more
confident level.


\vspace{1cm}

$Acknowledgements:$ We are grateful to E. Oset for the helpful comments and
proofreading, and would also like to thank Hao Chen for the careful
reading of the manuscript and helpful discussions. At last, this work
is supported in part by National Nature Science Foundations of China
under Contract Numbers  12335002, 12375078, 11975028. H.Q. Zheng and Z. Xiao are also
supported by ``the Fundamental Research Funds for the Central
Universities''.
\renewcommand\refname{Reference}
\bibliographystyle{h-physrev}
\bibliography{tcc}
\end{document}